# Growth Engineering of SrNbO$_3$ Perovskite Oxide by Pulsed Laser Deposition and Molecular Beam Epitaxy

*Jasnamol Palakkal, * $^{§, †}$ Tobias Meyer, $^{§}$ Márton Major, $^{†}$ Laurin Elias Bierschenk, $^{†}$ Lambert Alff $^{†}$*

$^{§}$Institute of Materials Physics, Georg-August-University of Göttingen, 37077 Göttingen, Germany

$^{†}$Institute of Materials Science, Technical University of Darmstadt, 64287 Darmstadt, Germany

*Email: jpalakkal@uni-goettingen.de

**ABSTRACT:** Molecular beam epitaxy (MBE) is a state-of-the-art technique for depositing thin films with precise stoichiometric control. However, when depositing oxides of perovskite-type ABO$_3$, this process becomes challenging as controlling the flux rate of A and B simultaneously in the presence of oxygen is difficult. In this work, by utilizing e-beam-assisted oxide MBE, we successfully deposited SrNbO$_3$ epitaxial thin films. A buffer layer of SrTiO$_3$ perovskite oxide, prepared by pulsed laser deposition (PLD), was used to improve the growth of this oxide significantly. This method overcomes the ultimatum of stabilizing perovskite oxide growth using MBE since the valence mismatch between the substrate and the films were effectively masked by the ultra-thin SrTiO$_3$ and further supported interfacial charge transfer. Interestingly, we also discovered that the perovskite oxides SrNbO$_3$ and SrTiO$_3$ replace K in KTaO$_3$ substrates by Sr, regardless of the deposition technique used, which is generally considered a reason for many interfacial effects. The growth of SrNbO$_3$ was unaffected by the K deficiency when the stable buffer layer was introduced.

## INTRODUCTION

The interface to the substrate is an inevitable component of a thin film system. Epitaxial stabilization of a material aiming to modify its functional properties often partners with fascinating novel functionalities. Defects introduced and alterations in crystal and electronic structures are critical in materials engineering using thin film technology. Additionally, interface and surface effects play an essential role, especially in epitaxial thin films locked on a substrate. During ultra-high vacuum (UHV) based physical vapor deposition (PVD) of oxides, the substrates are usually heated to support phase formation and achieve good crystalline quality of the films. The heated substrate, when in contact with the vapors of the material, can trigger different transport of the vapor. This includes various thermodynamic and kinetic means, which decide the nucleation and growth of the new thin-film crystal on the substrate. [1] Along with adsorption and desorption of particles, surface diffusion can also occur during this process. [1]

Substrate/lattice mismatch often introduces stress and strain to the lattice with surface atomic reconstruction. Typically, this happens only at the film lattice (prominent when the film is thermodynamically less stable) [2,3] since the kinetics hinders defect formation within the thick substrate. However, the substrate lattice can accommodate diffused atoms and charged particles of the vapor. Surface conductivity and polarity differences between substrate and film can also lead to interfacial reconstruction. [2,4] The interfacial energy also decides the growth mode of the film. [4] Nakagawa *et al.* reported that the atomic reconstruction at the interface could be avoided in multivalent oxides by allowing a charge flow across the substrate/film interface. [2] A charge transfer occurs at stable interfaces of oxides until a band alignment is reached since the *p* states of the oxygen network thrive to be continuous across the interface. [5] This implies that an interface with a proper charge transfer offers an atomically distinct, sharp interface. [2,5]

Molecular beam epitaxy (MBE), which utilizes individual elemental evaporation sources in the form of effusion cells and e-beam evaporation, is a primary tool used in the semiconductor industry due to its ability to fine-dope elements as per requirements. MBE offers a low average energy per incoming vapor particle compared to other UHV techniques, such as pulsed laser deposition (PLD) and ion beam-assisted deposition. [6] However, increased energy of the vapor favors the mobility of the atoms and, thus, a fast-homogeneous phase formation on the surface of the substrate. [6,7] Moreover, surface diffusion is normal during MBE growth processes. [8,9] A substrate rotation is usually implemented in MBE processes to ensure a homogeneous film growth and to avoid surface diffusion leading to lateral epitaxy during nucleation. [10]

In our work, we fabricated SrNbO$_3$, a transparent conducting material, [11, 12] using MBE, with a long-term goal of modifying its plasma frequency by varying the Sr/Nb ratio since it is convenient to use an MBE technique with elemental sources for precise controlling of the stoichiometry. Various resistivity values of ~28 to ~970 μΩ cm [11-16] were reported to be exhibited by SrNbO$_3$ thin films, fabricated by single-source pulsed laser deposition (PLD) and sputter deposition, irrespective of the choice of substrate but depending upon the defect concentrations. These works [11-17] used pseudo-cubic substrates, (LaAlO$_3$)$_{0.3}$-(Sr$_2$AlTaO$_6$)$_{0.7}$ (001) (LSAT), SrTiO$_3$ (001) (STO), DyScO$_3$ (110) (DSO), GdScO$_3$ (110) (GSO), KTaO$_3$ (001) (KTO), of different lattice mismatches with SrNbO$_3$ (lattice constant, c = 4.023 Å). [13] The resistivity values reported in recent literature are plotted as a function of pseudocubic lattice constant ($a_{pc}$) of the used substrates in **Figure 1a**.[11-17] The formation of uncontrolled defect concentration during the thin film growth including slight deviation in the stoichiometry can cause scattering and change the physical properties of perovskite epitaxial thin films, including resistivity.

It is known that thin films deposited from stoichiometric PLD targets can produce non-stoichiometric thin films due to different laser ablation thresholds of the elements. Due to this reason, in this work, we focused on the growth of SrNbO$_3$ thin films by oxide MBE by co-evaporation of elemental sources Sr and Nb along with a supply of molecular oxygen. To our knowledge, this work is the first attempt to grow SrNbO$_3$ films by oxide MBE. Recently, SrNbO$_3$ films were reported to be grown by using Sr and tris(diethalamido)(tert-butylimido) Nb organometallic precursor sources with a chemical vapor deposition (CVD) technique, conveniently known as 'hybrid-MBE' to check the surface stability of metastable SrNbO$_3$ films grown by hybrid-MBE. [18] The physical properties, such as electrical and optical properties, were not reported in their work. [18]

We used KTO (c = 3.989 Å) and GSO (3.965 Å) substrates in this project. KTO with cubic crystal structure was used due to low lattice mismatch with SrNbO$_3$; nevertheless, it was reported to show volatility at high temperatures. [19] Cubic films grown on orthorhombic GSO substrates (with pseudo-cubic (110) orientations) tend to show increased interfacial defects; further, they show electronic reconstruction at the interface. [3] Eu$^{2+}$Mo$^{4+}$O$_3$ films grown directly on Gd$^{3+}$Sc$^{3+}$O$_3$ (110) showed no film peak since the interface is electronically unfavorable for interfacial charge transfer. [3] This can be avoided by using a buffer layer of Sr$^{2+}$Ti$^{4+}$O$_3$ since Sr$^{2+}$Ti$^{4+}$O$_3$/Gd$^{3+}$Sc$^{3+}$O$_3$ interface can overcome such an electronic reconstruction due to the high thermodynamic stability of SrTiO$_3$ compared to EuMoO$_3$; at the same time, Sr$^{2+}$Ti$^{4+}$O$_3$/ Eu$^{2+}$Mo$^{4+}$O$_3$ interface is electronically favorable. [3] SrMoO$_3$, another transparent conducting oxide, grown by MBE on SrTiO$_3$ buffered KTO and SrTiO$_3$ buffered STO substrates, showed better film quality and thus better conductivity values than when deposited without a buffer layer. [20] It is not clear why the STO substrates need to be buffered with SrTiO$_3$ films itself, but it can be assumed due to the conductivity difference between SrMoO$_3$ film (metallic) and STO substrate (insulator). [20] At the same time, a few monolayers of SrTiO$_3$ buffer can be conducting due to defects. SrMoO$_3$ on Nb-doped STO, a metallic substrate, was also reported but did not address the difference in the film quality when deposited on SrTiO$_3$ buffered and non-buffered Nb-doped STO substrate. [20]

The primary aim of this work is to establish the growth of SrNbO$_3$ by using co-evaporation of elemental sources of oxide MBE. We deposited SrNbO$_3$ using MBE on non-buffered GSO and KTO as well as on substrates with a SrTiO$_3$ buffer layer deposited by PLD to improve the quality of the films further. This manuscript addresses the growth of SrNbO$_3$ and different interfaces observed in these perovskite films by means of standard materials characterization.

**EXPERIMENTAL DETAILS**

**Thin film deposition**. We prepared SrNbO$_3$ thin films using a custom-made oxide MBE setup. [21] E-beam evaporation was employed for the evaporation of Sr and Nb; molecular oxygen was used as the oxygen source. We used Sr cylinder and Nb pellets (Kurt J. Lesker Company®, 99.95% purity) and kept them in FABMATE® crucible inserts with a copper crucible liner during the evaporation. The photographs of the prepared Sr and Nb sources taken before loading into the load-lock of the source carousel chamber are shown in supplementary **Figures S1a** and **S1b**, respectively. The Sr and Nb elemental sources were kept in two of the six e-gun pockets that our MBE is equipped with. [21] Molecular oxygen was supplied via a leak valve aimed at the substrate. A high-power e-beam was used to evaporate the elements to produce the required rate of the Sr and Nb beam flux. Before each deposition, the Nb pellets were made into a single source block by melting them using the e-beam source. The Sr source was cleaned to remove any surface oxides by heating it above the vaporization temperature of Sr using the e-beam. By varying the energy of the e-beam, we configured various flux rates of Sr and Nb. Quartz crystal microbalances (QCM) were used to calibrate and monitor the flux rate of Sr and Nb. The substrate was heated to the growth temperature using a diode laser. The temperature from the front and back sides of the substrate was monitored by a two-color pyrometer. The substrate was continuously rotated during the growth using a motorized substrate manipulator to ensure the homogeneity of the films. A substrate shutter in front of the substrate manipulator and a growth chamber shutter in the middle of the chamber prevented the substrate from getting contaminated until the required rate for each elemental flux was reached.

We used Sr and Nb constant flux rates of 0.311 and 0.1 Å/s, respectively, during the entire growth duration. An oxygen flow of 0.2 sccm was maintained during the growth, which increased the base pressure of the growth chamber from ~$1\times10^{-11}$ to ~$1\times10^{-6}$ mbar. The substrates KTO (100) and GSO (110) were heated to 600 ºC and kept there for 30 minutes to remove any surface oxide. They were then cooled down to 500 ºC for the deposition. The heating and cooling rates were 50 and 30 ºC per minute, respectively. The substrates were rotated with a speed of 0.2 rpm during the deposition to ensure homogeneous film growth and to reduce surface diffusion.

The SrTiO$_3$ buffer layers (4 - 5 monolayer thick) were grown using PLD (Coherent Compex 205 KrF excimer laser, 248 nm) from a stoichiometric target. We used a laser frequency of 2 Hz and a fluence of 0.6 Jcm$^{-2}$. The deposition temperature was 630 ºC. The growth was performed with no additional oxygen flow under UHV conditions at the base pressure of ~$1\times10^{-8}$ mbar. After the deposition of STO, they were transferred to the MBE growth chamber through UHV tunnels without exposure to the air. Once they reached the MBE chamber, the temperature was set to 500 ºC, and then the SrNbO$_3$ film deposition continued. After the deposition, a 1 nm thick Pt protective coating was made on SrNbO$_3$ films at room temperature. A flux rate of 0.06 Å/s was used for Pt deposition. A room-temperature (cold) deposition was employed to avoid diffusion of Pt atoms to the SrNbO$_3$ lattice.

Reflection high-energy electron diffraction (RHEED) was employed before and after the deposition to get the diffraction patterns. A voltage of 49.9 kV and an emission current of 1.4 mA were used. We used kSA 400 to analyze the RHEED patterns. Using the RHEED photodetector during the growth was difficult due to the very bright light inside the chamber emitted during the e-beam evaporation of Nb.

**X-ray diffraction (XRD) and X-ray photoelectron spectroscopy (XPS).** Rigaku Smartlab® diffractometer in a parallel beam geometry using a Ge(220)×2 monochromator with Cu K$\alpha$ radiation was used for the X-ray diffraction (XRD) general 2θ scan, X-ray reflectivity (XRR), and reciprocal space mapping (RSM) measurements. The X-ray photoelectron spectroscopy (XPS) measurements were performed using a PHI Versaprobe 5000 spectrometer with monochromatic Al Kα radiation.

**Scanning transmission electron microscopy (STEM) imaging and electron energy loss spectroscopy (EELS).** Scanning transmission electron microscopy (STEM) specimens were prepared in a ThermoFisher Scientific Helios G4 dual beam focused ion beam using an acceleration voltage of 2 kV during the final Ga ion thinning step. The STEM and electron energy loss spectroscopy (EELS) analysis were carried out in an FEI Titan 80-300 E-TEM using a beam current of 30 pA and a convergence semi-angle of 10 mrad. The inner and outer collection semi-angles of the annular dark-field (ADF) detector were 47 mrad and 90 mrad, respectively, and the dispersion on the Ultrascan 1000XP camera included in the GIF Quantum 965 ER employed during EELS acquisition was 1 eV/channel. EELS line profiles were analyzed using the multiple linear least squares (MLLS) routine implemented in Gatan DigitalMicrograph (version 3.53.4137.0) to extract relative changes in the Ta M-, Sr L- and K K-edge intensities in growth direction.

## RESULTS AND DISCUSSION

**Thin film growth.** The thermodynamic stability window of the $Sr^{2+}$ and $Nb^{4+}$ under various oxygen partial pressures is critical while growing $SrNbO_3$ by co-evaporation of elements. The phase diagram of the oxidation of Nb metal into $Nb^{2+}$, $Nb^{4+}$, and $Nb^{5+}$ is shown in **Figure 1b** over the range of temperatures 400 ºC - 800 ºC that are feasible in our MBE setup. [22] The red-shaded area between the red and black lines in **Figure 1b** illustrates the stability of $Nb^{4+}$. Sr is stable in a +2 oxidation state for the full range of temperatures and equilibrium $pO_2$ presented in the plot. [22] This shows that there is enough growth window for $SrNbO_3$ by co-evaporation of Sr and Nb elemental sources.

The RHEED patterns captured before and after the $SrNbO_3$ deposition (at the growth temperature of 500 ºC) are shown in Supplementary **Figure S2**. The RHEED before deposition shows clear streaks indicating flat single crystalline surfaces with small domains. [23] The modulated streaks in the RHEED after deposition portray a multilevel stepped surface. [23] Compared to other samples, the films on KTO exhibit a rough surface.

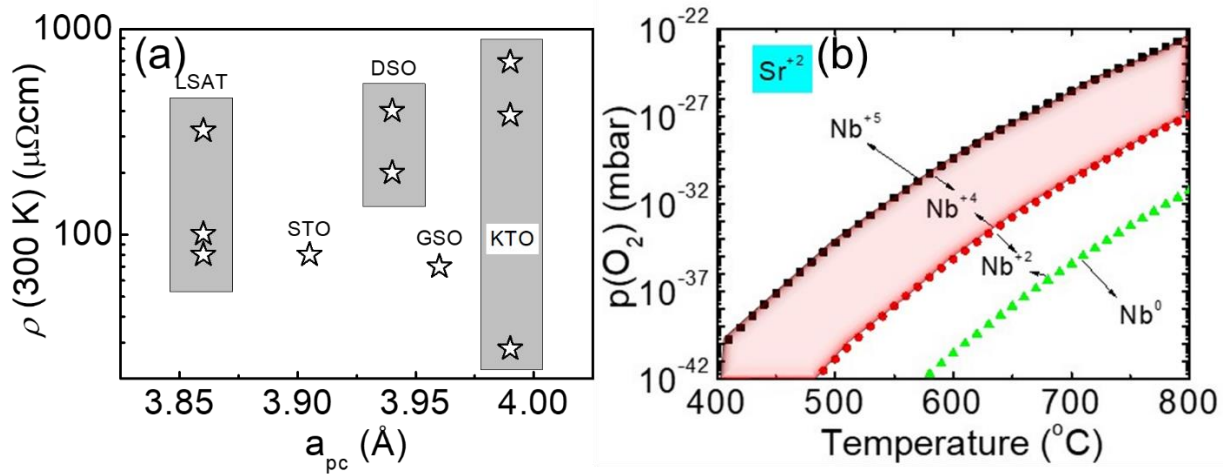

**Figure 1** (a) Resistivity of the SrNbO$_3$ epitaxial thin films on different substrates plotted as a function of the pseudocubic lattice constant of the perovskite substrates (LaAlO$_3$)$_{0.3}$(Sr$_2$AlTaO$_6$)$_{0.7}$ (LSAT), SrTiO$_3$ (STO), DyScO$_3$ (DSO), GdScO$_3$ (GSO) and KTaO$_3$ (KTO), collected from recent literature. [11-17] (b) The thermodynamic phase diagram of the oxidation of Nb. The red shaded area represents the stability window of Nb$^{4+}$. Sr$^{2+}$ is stable in the full range temperature and equilibrium $p$O$_2$ displayed in the plot.

**X-ray diffraction and X-ray photoemission spectroscopy.** The X-ray diffraction (XRD) 2θ scan patterns of SrNbO$_3$ films grown on GSO and KTO are shown in **Figures 2a** and **2b**, respectively. These patterns for a wide 2θ range are shown in supplementary **Figure S3**, which shows that the films are epitaxially grown with (00$l$) orientation. The SrNbO$_3$ film grown on GSO shows a weak reflection corresponding to (002) orientation. Upon inserting the buffer layer, the film quality improved, which can be seen as the presence of Laue oscillations, indicating a coherent crystalline quality. For the case of KTO, though the lattice mismatch is smaller than that with GSO, we didn't observe any film peak. With the SrTiO$_3$ buffer layer, the SrNbO$_3$ phase with (002) reflection can be seen in **Figure 2b**, along with Laue oscillations as in the case of SrNbO$_3$ buffered GSO substrates. As discussed earlier, providing an adequate interface is necessary for proper thin film epitaxial growth. Here, we have a valence-mismatched (pseudo)cubic substrate and film (Gd$^{3+}$Sc$^{3+}$O$_3$/Sr$^{2+}$Nb$^{4+}$O$_3$ and K$^{1+}$Ta$^{5+}$O$_3$/Sr$^{2+}$Nb$^{4+}$O$_3$). A few monolayers of STO hindered this electronic mismatch and provided a sharp interface. [2, 3] The out-of-plane lattice constant ($c$) is calculated to be 4.061 Å for GSO, 4.079 Å for GSO+STO, and 4.084 Å for KTO+STO.

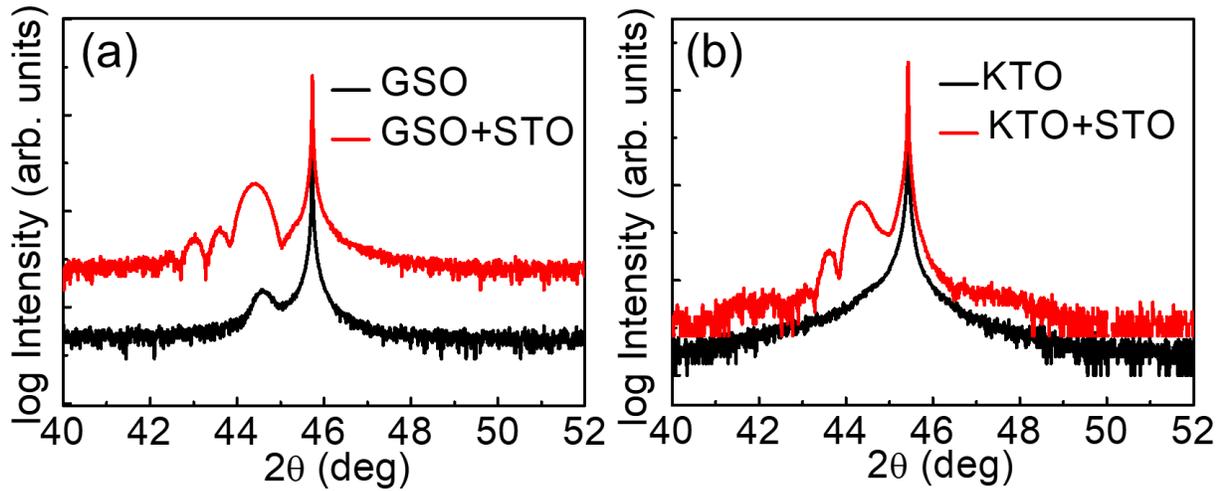

**Figure 2.** XRD pattern of SrNbO$_3$ epitaxial thin films grown (a) on GdScO$_3$ (110) GSO substrate with and without SrTiO$_3$ (STO) buffer layer and (b) on KTaO$_3$ (001) (KTO) substrate with and without STO buffer layer.

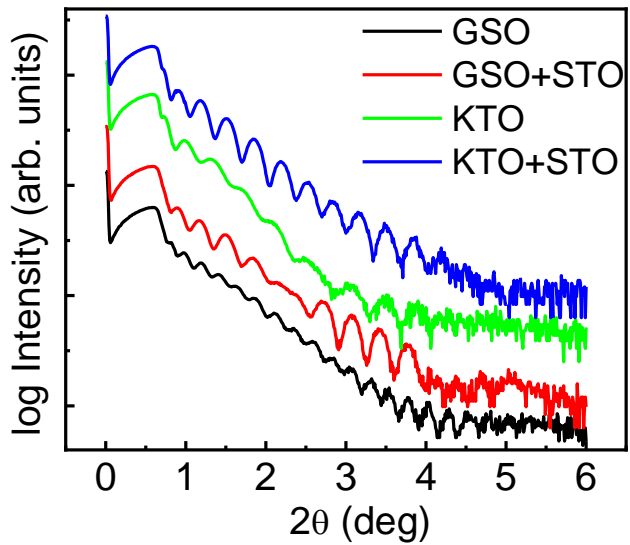

**Figure 3.** The XRR patterns of SrNbO$_3$ films.

**Figure 3** shows the films' X-ray reflectivity (XRR) patterns. All the films show oscillations in the XRR corresponding to the Kiessig thickness-fringes, portraying a relatively smooth substrate/film interface. However, the slope changes in the XRR at higher 2θ angles hint at surface and interface roughness in these films.

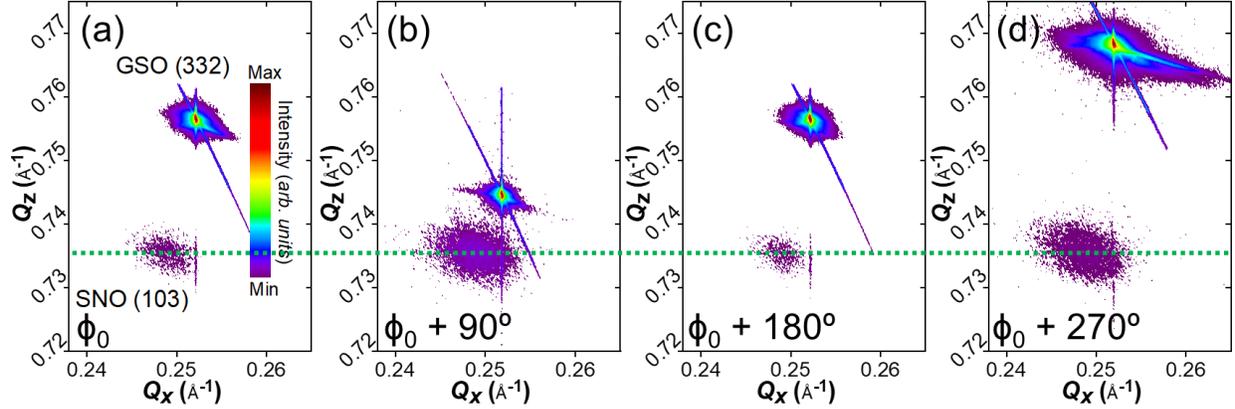

**Figure 4.** Reciprocal space mapping (RSM) of SrNbO$_3$ films grown on GSO+STO substrate with azimuthal angle **(a)** $\Phi_0$, **(b)** $\Phi_0$ + 90º, **(c)** $\Phi_0$ + 180º, and **(d)** $\Phi_0$ + 270º.

A reciprocal space mapping (RSM) was performed to check how much the films are strained to the substrate. In search of the SrNbO$_3$ (103) peak, we measured around the GSO (332) and KTO (103) reflections. To check the SrNbO$_3$ film growth on orthorhombic GSO substrate, we measured the RSM in four azimuthal (Φ) angles, with an increment of 90 deg, as shown in **Figure 4a-d**. The film has the same out-of-plane lattice constant $c$ = 4.08 Å in all four azimuthal angles, as marked by the green dashed line. This shows that SrNbO$_3$ has grown perpendicular to the surface of GSO (which is pseudocubic, but the actual reciprocal lattice is tilted with respect to the substrate's surface). SNO only took the in-plane constraint and did not follow the orthorhombic GSO lattice. The value of $c$ is the same as that measured from the XRD 2θ scan. However, the in-plane film quality is not uniform, and the film is not fully locked to the substrate. The in-plane lattice constant, $a$ = 4.008 Å, is slightly off from that of the substrate. The RSM image of the film on KTO+STO is shown in **Figure 5a**. Unlike the former film, SrNbO$_3$ on KTO+STO is better in-plane locked to the substrate's lattice. This could be due to the less lattice mismatch between SrNbO$_3$ and KTO. We calculated the in-plane strain ($\varepsilon_a$) and out-of-plane strain ($\varepsilon_c$) by taking the bulk lattice constant of SrNbO$_3$ as 4.023 Å.[24, 25] SrNbO$_3$ on GSO+STO has $\varepsilon_a$ = -2.79 % and $\varepsilon_c$ = 1.39 % with respect to the bulk lattice. SrNbO$_3$ on KTO+STO has $\varepsilon_a$ = -2.33 % and $\varepsilon_c$ = 1.52 %. Compared with a bulk SrNbO$_3$ crystal, SrNbO$_3$ films on GSO+STO are more in-plane strained than those on KTO+STO.

On the other way, in the out-of-plane direction, the film on KTO+STO is more relaxed than the film on GSO+STO. This is due to the fact that the films on KTO+STO are locked in-plane to the substrate, and the total volume of the unit cell is adjusted by elongating in an out-of-plane direction to retain the volume of the bulk crystal.

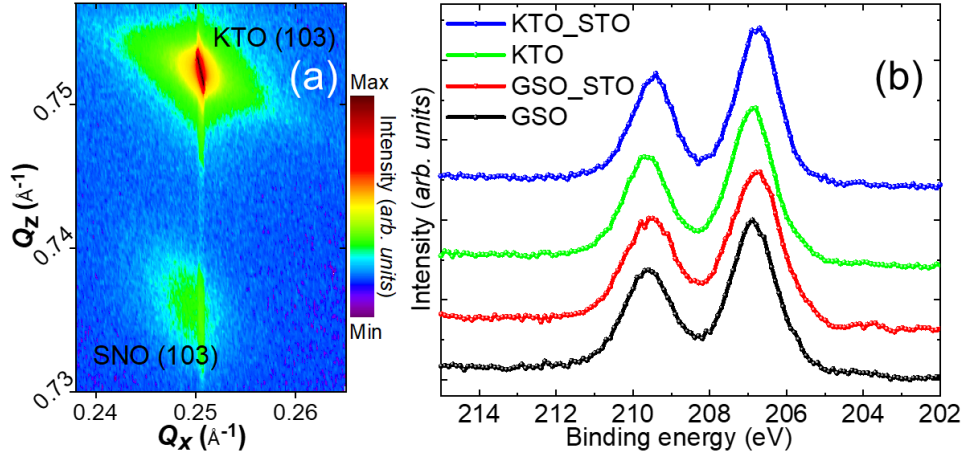

**Figure 5.** **(a)** Reciprocal space mapping (RSM) of SrNbO$_3$ films grown on KTO+STO substrate. **(b)** Nb 3$d$ XPS spectra of SrNbO$_3$ films.

To check the oxidation state of Nb in the deposited SrNbO$_3$ films, we measured the Nb 3$d$ XPS spectra, as shown in **Figure 5b**. We observed the same spectra for all the samples, with the Nb$^{4+}$ 3$d_{5/2}$ peak appearing at ~206.8 eV. The satellite peak originating from the final-state effect of electrons, as observed in metallic/semi-conducting Nb-oxides, [26] is absent here. Since this satellite peak is relatively weak, the Pt metallic coating could have hidden the significant signals originating from the final-state effect.

**Scanning transmission electron microscopy (STEM) with electron energy loss spectroscopy (EELS).**

The STEM images in **Figure 6** reveal the microstructure of the SrNbO$_3$ films grown on (a) GSO, (b) GSO+STO, (c) KTO, and (d) KTO+STO. The red arrows display the substrate/SrNbO$_3$ and substrate/STO interfaces and the yellow arrows show the STO/SrNbO$_3$ interfaces. The SrNbO$_3$ films grown directly on the respective substrates are strongly disordered and thus show only weak atomic contrasts. However, the coherently grown SrTiO$_3$ buffer layer promotes a higher crystalline quality of the SrNbO$_3$ film in accordance with the XRD results presented in **Figure 2**.

Nevertheless, even with the SrTiO₃ buffer layer, a relatively high concentration of planar defects is observed in the SrNbO₃ films. In addition, both samples with a KTO substrate exhibit a bright layer at the interface between KTO and SrNbO₃, as well as between KTO and SrTiO₃.

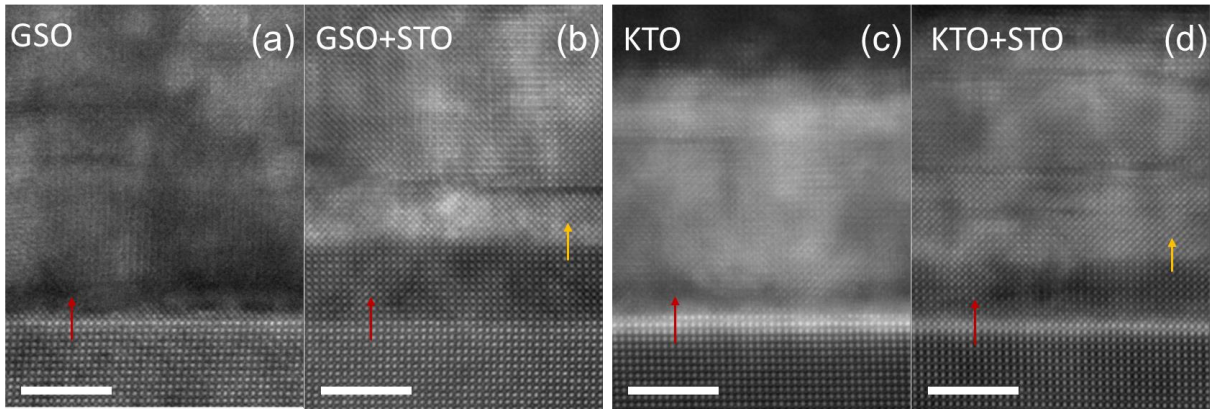

**Figure 6.** Atomically resolved ADF-STEM images of SrNbO₃ films grown on **(a)** GSO, **(b)** GSO+STO, **(c)** KTO, and **(d)** KTO+STO substrates. In (c) and (d), the Pt protection bar of the FIB preparation process is visible in the top part of the images due to a thinner SrNbO₃ film compared to (a) and (b). The red and yellow arrows are drawn at the substrate (KTO or GSO) to film (STO or SrNbO₃) and STO to SrNbO₃ interfaces, respectively. The scale bar is 5 nm.

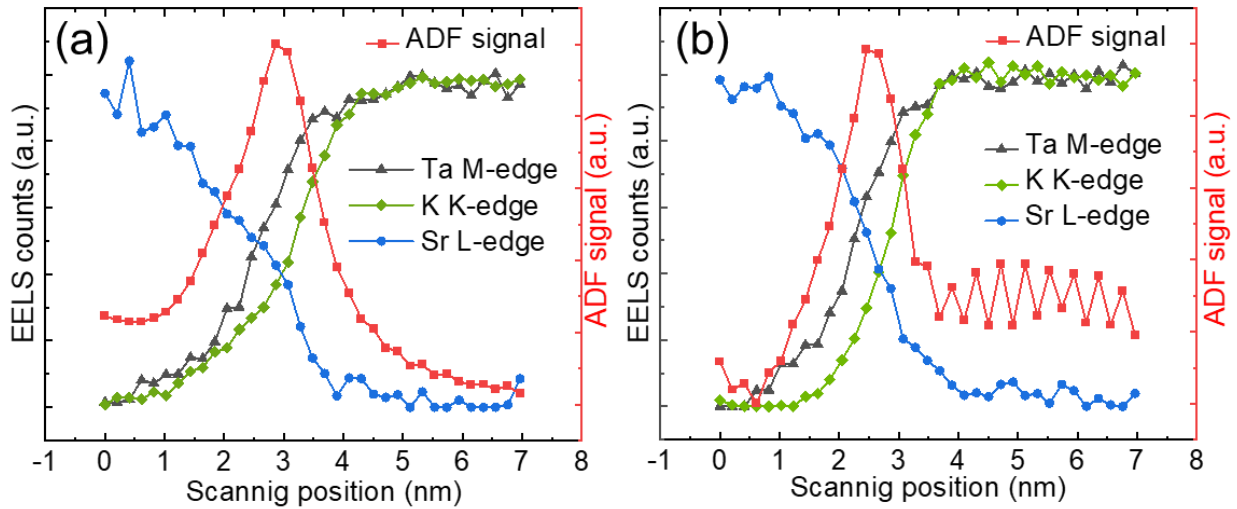

**Figure 7.** EELS analysis of the interfaces to KTO: **(a)** Shows the relative intensity of the Ta M-, Sr L-, and K K-edge with respect to the position in the growth direction for SrNbO₃ growth on KTO. **(b)** Same as in (a) but for growth of SrNbO₃ on KTO+STO.

In order to analyze chemical heterogeneities as possible origins for the bright layer, EELS profiles of the Ta M-, Sr L-, and K K-edge intensities are shown in **Figure 7** for the KTO samples with and without STO buffer. The results reveal a decreased potassium signal in the bright layer, suggesting its substitution by Sr in the top part of the KTO substrate. Similar replacements of K with other elements were reported in KTO interfaces. LaAlO$_3$ and EuO deposited on KTO substrates replaced the K with La and Eu, respectively, causing functionalities like two-dimensional superconductivity and anisotropic transport. [27] In our work, the perovskite SrTiO$_3$ (PLD-grown) and SrNbO$_3$ (MBE-grown) replaced the K with Sr. Irrespective of the technique used for the growth of these two perovskite oxide films, the K was replaced.

The SrNbO$_3$ film grown by MBE directly on GSO provides a good interface, where one to two monolayers are grown before the film starts showing defects, as seen in **Figure 6a**. In the case of the film on KTO, the film is primarily defective, and no sharp interface can be seen (**Figure 6c**), which is in agreement with the XRD results. Local nanocrystalline regions are visible in both SrNbO$_3$/GSO and SrNbO$_3$/KTO films, but the atomic replacement at the KTO substrate leads to poor thin film growth even though KTO is more lattice-matched with SrNbO$_3$. Growth of oxide perovskites by MBE is a complex process as one has to control the rate of both A and B species together with the flow of oxygen. Though Sr has replaced K, SrTiO$_3$ grows epitaxially on KTO (**Figure 6d**) since the increased energy of vapors during the PLD process promotes fast film growth. The PLD vapors usually have considerable kinetic energy (hundreds of eV), and MBE vapors typically have as low as less than one eV. [28] Later on, the SrNbO$_3$ grows very well on top of this SrTiO$_3$ buffer layer (**Figure 6d**). It is also worth mentioning the conductivity difference at the interface. Both KTO and GSO are insulators, while SrNbO$_3$ is a metallic conducting oxide. Thin films of SrTiO$_3$ are reported to exhibit low conductivity due to defect states. [29] The valence mismatch and conductivity difference across the GSO/ SrNbO$_3$ and KTO/ SrNbO$_3$ interfaces were hindered by the usage of SrTiO$_3$ thin films, which made the charge transfer across the interface plausible. Previous works used SrTiO$_3$ buffer layers even when using STO substrates. [20]

**CONCLUSION**

Perovskite oxide SrNbO$_3$ was grown on a GSO substrate by co-evaporation of elemental sources using an oxide MBE. Defects prevail in this film, where only a few monolayers are immaculate. The growth of SrNbO$_3$ by MBE was further improved by adding a buffer layer of SrTiO$_3$ grown by PLD. SrNbO$_3$ perovskite growth was confirmed on the orthorhombic GSO. Though the cubic KTO substrate offered a good lattice match with SrNbO$_3$, a better epitaxial thin film growth was not achieved on KTO. However, adding a SrTiO$_3$ buffer layer stabilized the growth of SrNbO$_3$ on KTO. We found that at the KTO interface, the K elements were replaced by Sr when both SrNbO$_3$ and SrTiO$_3$ were grown, irrespective of the growth technique. The fast growth of thermodynamically stable SrTiO$_3$ by PLD was unaffected by the K deficiency on KTO, which aided the charge transfer across the interface and made the SrNbO$_3$ growth by MBE feasible. Using MBE to change the stoichiometry for tuning functional properties is challenging when the interface has defects and is not electronically matched. Our work shows this hassle can be overcome using a buffer layer deposited by a fast process like PLD.

ASSOCIATED CONTENT

**Supporting Information**.

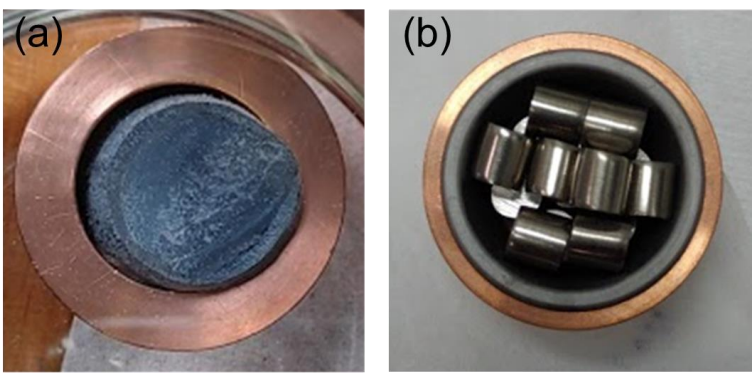

**Figure S1.** The photographs of the prepared (a) Sr and (b) Nb sources made before loading into the load-lock of the source carousel chamber of MBE.

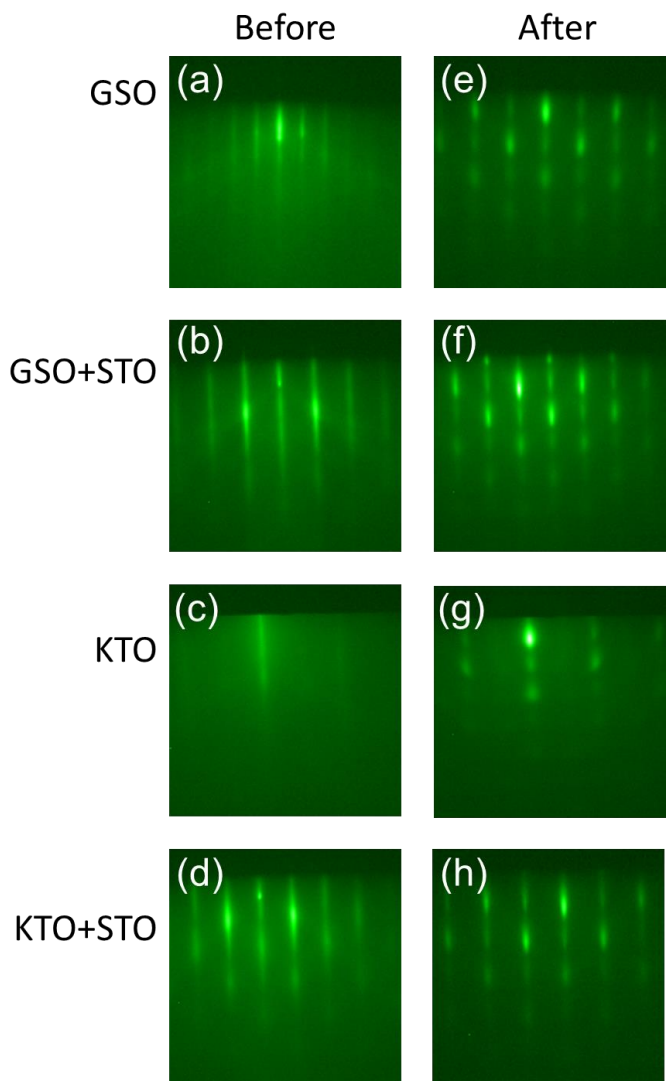

**Figure S2.** RHEED patterns taken before and after SrNbO$_3$ thin film deposition.

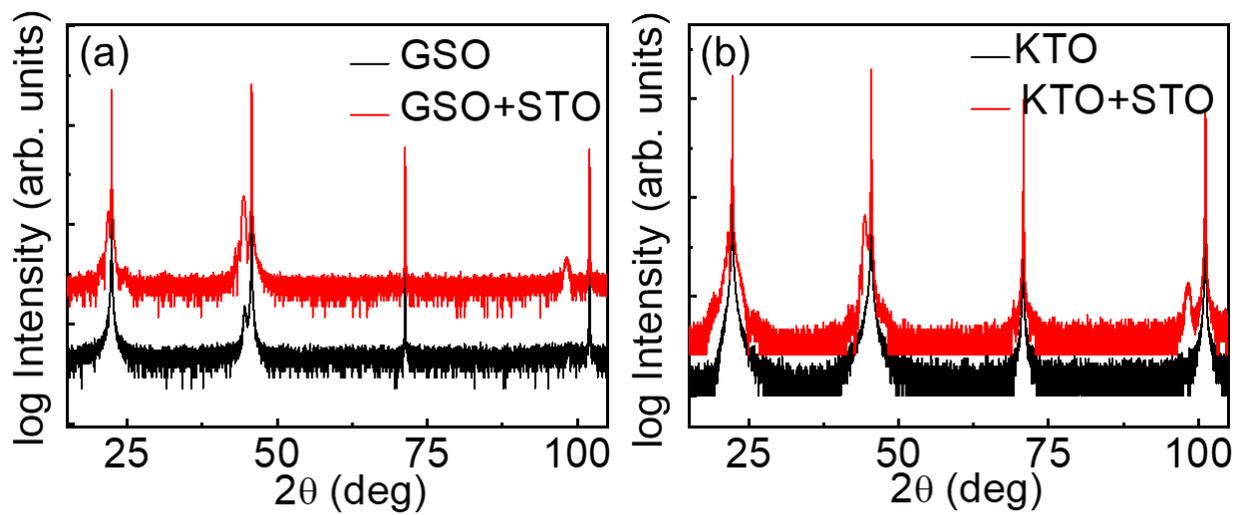

**Figure S3.** XRD pattern (for a 2θ range of 15 to 105 deg) of SrNbO$_3$ epitaxial thin films grown **(a)** on GdScO$_3$ (110) GSO substrate with and without SrTiO$_3$ (STO) buffer layer and **(b)** on KTaO$_3$ (001) (KTO) substrate with and without STO buffer layer.

AUTHOR INFORMATION

**Corresponding Author**

Jasnamol Palakkal − Institute of Materials Physics, Georg-August-University of Göttingen, 37077 Göttingen, Germany. Institute of Materials Science, Technical University of Darmstadt, 64287 Darmstadt, Germany. Email: jpalakkal@uni-goettingen.de

**Authors**

Tobias Meyer − Institute of Materials Physics, Georg-August-University of Göttingen, 37077 Göttingen, Germany

Márton Major, Laurin Elias Bierschenk, Lambert Alff − Institute of Materials Science, Technical University of Darmstadt, 64287 Darmstadt, Germany

**Author Contributions**

J.P. - Designed and performed the experiments. Collected and analyzed the data. Wrote and edited the manuscript. Funding acquisition.

T.M. - Performed the experiments. Collected and analyzed the data. Wrote and edited the manuscript.

M.M. - Assisted with the experiments, data collection, and analysis. Edited the manuscript.

L.E.B. - Performed the experiments and data collection.

L.A. - Designed the experiments. Wrote and edited the manuscript.

**Funding Sources**

Deutsche Forschungsgemeinschaft (DFG) projects 429646908 and 217133147/SFB 1073.

**Notes**

The authors declare no competing financial interest.

ACKNOWLEDGMENT

This work was supported by Deutsche Forschungsgemeinschaft (DFG, German Research Foundation) under projects 429646908 and 217133147/SFB 1073. The use of equipment in the "Collaborative Laboratory and User Facility for Electron Microscopy" (CLUE) at the University of Goettingen is gratefully acknowledged. We thank Patrick Salg and Lukas Zeinar, TU Darmstadt for contributing to the PLD deposition of STO layers. Gabriele Haindl, TU Darmstadt is greatly acknowledged for the technical help received during the oxide MBE usage.